\documentclass[10pt,conference]{IEEEtran}
\IEEEoverridecommandlockouts

\usepackage{cite}
\usepackage{amsmath,amssymb,amsfonts}
\usepackage{algorithmic}
\usepackage{graphicx}
\usepackage{textcomp}
\usepackage{xcolor}
\def\BibTeX{{\rm B\kern-.05em{\sc i\kern-.025em b}\kern-.08em
    T\kern-.1667em\lower.7ex\hbox{E}\kern-.125emX}}

\usepackage{multirow}

\usepackage[top=0.75in, bottom=1.25in, left=0.625in, right=0.625in]{geometry}

\usepackage{array}
\usepackage[caption=false,font=normalsize,labelfont=sf,textfont=sf]{subfig}
\usepackage{caption}
\usepackage{stfloats}
\usepackage{url}
\usepackage{verbatim}
\usepackage{needspace}
\usepackage{cite}
\usepackage{stfloats}
\usepackage[linesnumbered,ruled]{algorithm2e}
\newcommand{\rev}[1]{#1}

\usepackage{epstopdf}

\begin{document}
\title{\fontsize{23.8pt}{28pt}\selectfont Hardware-friendly Baseband Design and Performance Analysis for LEO Broadband LoS-MIMO Systems\\
\thanks{This work was supported in part Natural science Basic Research Project of Shaanxi Province under Grant 2024JC-YBMS-509, and in part the Youth Innovation Team of Shaanxi Universities. (Corresponding author: Fengkui Gong)
}
}

\author{Peixin Zhang$^{*,\dagger}$, Fengkui Gong$^*$, Guo Li$^\dagger$, and Tony Q. S. Quek$^\ddag$ \\
\footnotesize
$^*$ the State Key Laboratory of ISN, School of Telecommunications Engineering, Xidian University, Xi'an, 710071, China \\
$^\dagger$ the National Key Laboratory of Science and Technology on Space Microwave, China Academy of Space Technology (Xi’an), 710071, China \\
$^\ddag$ the Information System Technology and Design Pillar, Singapore University of Technology and Design, Singapore 487372\\
pxzhang@stu.xidian.edu.cn; fkgong@xidian.edu.cn; gli@xidian.edu.cn; tonyquek@sutd.edu.sg
}

\maketitle

\begin{abstract}
High-frequency and broadband line-of-sight multiple-input multiple-output (LoS-MIMO) is a key solution for massive space-to-Earth data downlinks. However, in low Earth orbit (LEO) scenarios, it faces spatial rank deficiency from time-frequency coupling correlation and clock deviations from distributed oscillators. This paper establishes a dynamic effective channel throughput (ECT) framework incorporating waterfall characteristics of forward error correction. A hardware-friendly OFDM-based transceiver architecture with subband-adaptive truncation and dynamic timing recovery to achieve robust spatial decoupling is proposed. Through rigorous closed-form derivations, the equivalent signal-to-interference-plus-noise ratio (SINR) under joint errors is quantified, and an optimal scheduling algorithm combining truncation with variable coding and modulation is proposed to maximize ECT. Simulations confirm that the design circumvents outages caused by spatial notches with low complexity, stably enhancing the potential average throughput of a single dual-polarized beam to over 100~Gbps, providing a reliable theoretical foundation and baseband solution for next-generation high-throughput satellites.

\end{abstract}

\begin{IEEEkeywords}
LEO satellite communications, LoS-MIMO, spatial rank deficiency, clock deviation, channel throughput analysis
\end{IEEEkeywords}

\section{Introduction}

\IEEEPARstart{T}{he} exponential growth of data generated by low Earth orbit (LEO) satellites has severely strained traditional Ka- and Ku-band data return links. To meet the imminent demand for Terabit-per-second (Tbps) downlinks, distributed line-of-sight multiple-input multiple-output (LoS-MIMO) operating in the E-band ({71--86~GHz}) has emerged as a promising solution. By employing distributed ground stations or formation-flying satellites, distributed LoS-MIMO achieves the large spatial separation required for full-rank multiplexing in pure LoS environments \cite{ref1, ref2}. However, pushing ultra-wideband distributed LoS-MIMO technology toward engineering applications still faces formidable challenges in baseband signal processing \cite{ref2-1,ref2-2}. Firstly, the deep coupling of continuous bandwidth over several GHz and large-scale distributed arrays induces high-frequency and non-uniform time-frequency coupling correlation, which is similar to the beam squint effects in massive MIMO \cite{ref3}. This effect, combined with deterministic orbital dynamics, inevitably causes severe spatial rank deficiency and system outage at specific geocentric angles. Secondly, the high-speed motion of LEO satellites and the use of independent local oscillators at physically isolated receiving nodes introduce severe Doppler shifts and sampling frequency deviations (SFD). \rev{The resulting unaligned fractional delays significantly complicate MIMO spatial decoupling} \cite{ref4}.

Currently, most system-level theoretical models for satellite MIMO still rely on statistical fading assumptions \cite{ref5}, frequently neglecting the fast time-frequency varying nature and broadband effects inherent to LEO pure LoS channels \cite{ref6, ref7}. Furthermore, performance evaluations universally adopt ideal Shannon continuous spectrum integration. The ideal continuous spectrum assumption masks various channel impairments in discrete baseband, easily leading to an overestimation of effective throughput. \rev{The mobility and heterogeneous resource coupling of space--air--ground networks also make repeated system-level simulation too costly for real-time link adaptation \cite{ref_sagin_survey}.} At the network-management layer, margin-aware relational learning has also been explored for imbalanced incremental fault diagnosis \cite{ref_fault_diag}. There remains a lack of an effective channel throughput (ECT) metric that incorporates micro-scale errors and practical forward error correction (FEC) coding thresholds. In practical system design, existing synchronization schemes mostly assume homologous clocks \cite{ref8}, failing to evaluate the destructiveness of micro-errors in distributed architectures. Moreover, state-of-the-art precoding algorithms designed to mitigate spatial rank deficiency such as graph neural networks (GNN) \cite{ref9} or Chebyshev-accelerated iterations \cite{ref10}, demand exceptionally high computational complexity. This makes them fundamentally difficult to implement on hardware like field-programmable gate arrays (FPGAs). 

This paper analyzes the impact mechanisms of spatial rank deficiency and clock deviation on broadband LoS-MIMO systems and proposes \rev{a hardware-friendly, robust} baseband processing architecture. The main contributions are summarized as follows:

\begin{itemize}
\item \rev{Based on a dynamic spatio-temporal coupling model of non-orthogonal and non-homologous LoS-MIMO channels, an ECT metric is established by combining pilot/CP overhead with the BLER characteristic obtained from CCSDS (8160,7136) LDPC decoding. Closed-form analysis then connects orbital geometry and residual SFD to demodulation error and useful throughput.}

\item \rev{A hardware-friendly OFDM transceiver with subband-adaptive truncation is developed. The associated finite-grid scheduler jointly selects the truncation threshold and square-QAM order and switches to a one-stream mode when spatial multiplexing becomes inefficient. Link-level simulations show an average single-polarization ECT of 55.98~Gbps over the evaluated angular interval.}
\end{itemize}  

The remainder of this paper is organized as follows: Section II establishes the channel model and introduces the ECT. Section III elaborates on the proposed baseband design. Section IV theoretically derives system performance and proposes the optimal scheduling algorithm. Section V verifies the architectures through simulations, and Section VI concludes the paper.

\textit{Notations}: $(\cdot)^T$, $(\cdot)^*$, and $(\cdot)^H$ denote transpose, complex conjugate, and conjugate transpose. $|\cdot|$ denotes the determinant of a matrix or absolute value. $\mathbf{X} \!\in\! \mathbb{C}^{a \times b}$ indicates a complex matrix of size $a \!\times\! b$. $\mathbf{I}_N$ represents an $N \!\times\! N$ identity matrix. $\text{Tr}(\mathbf{X})$ denotes the trace of $\mathbf{X}$. $\arg\{\cdot\}$ represents the phase angle of a complex number.

\vspace{0.5em}
\section{System Model and Equivalent Throughput} \label{sec2}

\subsection{Dynamic Channel Model}

Consider a LoS-MIMO downlink comprising a single-beam LEO satellite with an $N_t$-element uniform linear array (ULA)~\cite{s2} and $M_r$ distributed ground receiving stations forming a virtual ULA. The transmit and receive antenna spacings are $d_s$ and $d_g$, respectively. The satellite transmits with a total power $P$ equally distributed across the $N_t$ time-frequency homologous transmitters. 

\begin{figure}[t]
\vspace{0.5em}
\begin{center}
\includegraphics[width=0.99\columnwidth]{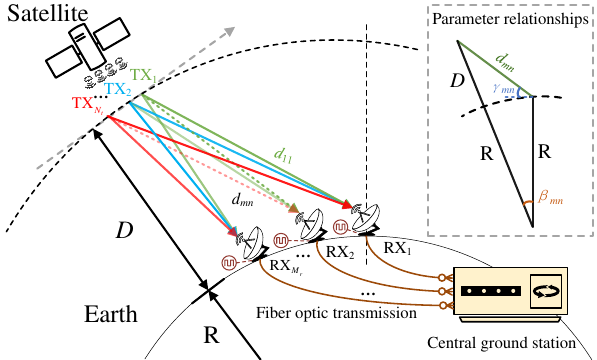}
\caption{Diagram of satellite-to-ground LoS-MIMO.}
\label{fig1}
\end{center}
\vspace{-0.5em}
\end{figure}

As shown in Fig.~\ref{fig1}, let $R$, $D$, and $\omega_{sat} = \sqrt{\mu/(R+D)^3}$ denote the Earth's radius, satellite orbit altitude, and angular velocity, respectively. At time $t$, given an initial geocentric angle $\theta_0$, the instantaneous geocentric angle $\beta_{m,n}(t)$ and propagation distance $d_{m,n}(t)$ between the $n$-th transmit and $m$-th receive antennas are given by:
\Needspace{5\baselineskip}
\begin{equation}
\!\beta_{m,n}(t)\!=\!\theta_0\!+\!\omega_{sat}t\!+\!\left(\!n\!-\!\!\frac{N_t\!+\!1}{2}\!\right)\!\frac{d_s}{R\!+\!D}\!\!-\!\left(\!m\!-\!\!\frac{M_r\!+\!1}{2}\!\right)\!\frac{d_g}{R},\!
\end{equation}
\begin{equation}
\!\!d_{m,n}(t)\!=\!\sqrt{(R+D)^2+R^2-2R(R+D)\cos\beta_{m,n}(t)}.
\end{equation}
The propagation delay is $\tau_{m,n}(t) = d_{m,n}(t)/c$. Assuming the $m$-th receiving station has an inherent local oscillator deviation ratio $\varpi_m$, the baseband equivalent channel frequency response around carrier $f_c$ over bandwidth $B$ is modeled as:
\begin{equation}
H_{m,n}(t,f) \!=\! e^{-j2\pi(f_c+f)\tau_{m,n}(t)}\!\cdot\! e^{-j2\pi \left(f_c\varpi_mt- f\tau_m^{bulk}(t)\right)}.
\end{equation}
where $\tau_m^{bulk}(t)$ represents the bulk delay of the $m$-th station. \rev{Ephemeris-aided coarse timing and a feedforward numerically controlled oscillator (NCO) remove the bulk geometric delay and deterministic common Doppler once per frame. Consequently, $\varpi_m$ represents the residual station-dependent sampling-frequency ratio after acquisition. The analytical model focuses on the SFD-induced interpolation error, while the remaining synchronization, estimation, quantization, ephemeris, and rain effects are evaluated through link-level simulation in Fig.~\ref{fig5}.}

\subsection{Theoretical Capacity and Effective Channel Throughput}

Assuming an ideal and oscillator-error-free channel matrix $\mathbf{H}_0(t,f)$, which is only composed of $\exp\{-j2\pi(f_c+f)\tau_{m,n}(t)\}$, the theoretical broadband MIMO capacity is
\begin{equation}
\!\!\!C_0(t)\!=\!\int_{-B/2}^{B/2}\!\!\log_2\left(\left|\mathbf{I}_{M_r}\!+\!\frac{P}{\sigma^2 N_t}\mathbf{H}_0(t,f)\mathbf{H}_0^H(t,f)\right|\right)df\!,\!
\end{equation}
where $P,\ \sigma^2$ is the total transmitted power and the power spectral density of the additive white Gaussian noise (AWGN), respectively. Distributed oscillator deviations introduce a unitary phase offset matrix $\boldsymbol{\Phi}(t,f)$, yielding the impaired channel $\mathbf{H}(t,f) = \boldsymbol{\Phi}(t,f)\mathbf{H}_0(t,f)$. Since $\boldsymbol{\Phi}(t,f)$ is unitary, the theoretical capacity remains strictly equal, meaning the theoretical spatial rank upper bound is unaffected by clock deviations.

{In practical systems, residual clock error and frequency-dependent rank loss increase the uncoded BER $P_{b,\mathrm{unc}}(t)$. Let $\Omega_v$ contain the $N_v$ retained subcarriers, and let $\eta_t=N_{\mathrm{data}}/(N_{\mathrm{pilot}}+N_{\mathrm{data}})$ and $\eta_c=7136/8160$ denote the time and coding efficiencies. The FEC waterfall is characterized by a BPSK/AWGN $E_b/N_0$ sweep of the CCSDS C2 code with normalized min-sum decoding and at most 30 iterations. With random interleaving across the retained tones, the frequency-selective uncoded BER is mapped to the fitted decoder transfer curve} $\widehat P_{\rm BLER}(p)\!=\!\left[1+\exp\{-64.26(\log_{10}p+1.597)\}\right]^{-1},$ {where $p=P_{b,\mathrm{unc}}$. The corresponding ECT is}
\begin{equation}
{\!\!C_{eq}(t)\!=\!R_{sym}N_t\frac{N_v(t)}{N_{\text{FFT}}}\eta_{\text{t}}\eta_{\text{c}}\log_2(M)\!\!\left[1\!-\!\widehat P_{\rm BLER}(P_{b,\rm unc})\right]\!.\!}
\end{equation}
The ECT serves as the core metric for evaluating the subsequent baseband designs.

\section{Proposed Baseband Transceiver Architecture and Subband-Adaptive Design}

\subsection{Subband-Adaptive Truncation Mechanism}

By utilizing high-precision LEO satellite ephemeris data, the satellite can extract prior transmission delay parameters to locate rank-deficient subcarriers. Given the instantaneous time $t$ and discrete frequency $f_k$, a theoretical prior channel matrix $\mathbf{H}_0(t,f_k)$ containing only spatial phase information is reconstructed. Iterating through all discrete subcarriers $f_k$ ($k \!\in\! [0,N_{\text{FFT}}\!-\!1]$) within the effective baseband, the prior condition number $\kappa(t,f_k) \!=\! \|\mathbf{H}_0(t,f_k)\| \!\cdot\! \|\mathbf{H}_0^{-1}(t,f_k)\|$ is calculated. A maximum condition number safety threshold $\kappa_{th}(t)$ tolerable by the system is set to generate a binary adaptive truncation mask vector $\mathbf{p}_v \!\in\! \{0, 1\}^{N_{\text{FFT}}}$, where $p_v(k) \!=\! 1$ if $\kappa(t,f_k) \!\le\! \kappa_{th}(t)$. The set of healthy subcarrier indices retained by the mask is $\Omega_v \!=\! \{k | p_v(k) \!=\! 1\}$, with its cardinality $N_v$. The method for optimizing set selection can be found in Sec.~\ref{sec4.3}.

\vspace{0.5em}
\subsection{Dynamic Interleaved Frame Structure and Pilot Design}

\rev{In LEO channels, residual phase evolution requires CSI updates over tens of symbols. Each frame therefore alternates $N_{\text{pilot}}=N_t$ orthogonal pilot symbols and $N_{\text{data}}$ data symbols. In the 4$\times$4 simulation, four pilot symbols and 20 data symbols give $\eta_t=20/24$, which is included in the ECT calculation.}

To achieve strict orthogonal decoupling of the $N_t$ transmit antennas, mutually coprime Zadoff-Chu (ZC) sequence roots are allocated in the FD for each antenna. The pilot base sequence for the $n$-th antenna is $Z_n(k) = \exp(-j\pi u_n k(k+1)/N_{\text{FFT}})$. On this basis, an $N_t$-th order Walsh-Hadamard orthogonal matrix $\mathbf{W} \in \{+1, -1\}^{N_t \times N_t}$ is introduced in the time domain to eliminate cross-talk caused by the parallel transmission of multiple antennas. In the $p$-th pilot symbol period ($p\in[1,N_t]$), the pilot transmitted by the $n$-th antenna is $W_{n,p} Z_n(k)$. Finally, the frequency-domain equalization requires a cyclic prefix (CP) to prevent aliasing. The CP length should cover the maximum delay difference of the spatial array after alignment to the reference time:
\begin{equation}
N_{\text{CP}} \ge \lceil R_{sym} \max_m \{ \max_n\{\tau_{m,n}(t)\} - \min_n\{\tau_{m,n}(t)\} \} \rceil.
\end{equation}

\vspace{0.5em}
\subsection{Channel Estimation and MMSE Equalization}

The observation signal captured by the $m$-th receive antenna at the $p$-th pilot symbol period and the $k$-th subcarrier is coherently accumulated over the $N_t$ continuous TD symbols, by comparing the ZC sequence of that antenna, it yields
\Needspace{6\baselineskip}
\begin{equation}
\!\!\!\!\!{\hat H_{m,{n_0}}}(k) \!=\!\! \frac{{\sum\limits_{n = 1}^{{N_t}}\!\! {\left[ {{H_{m,n}}(k){Z_n}(k)\!\!\left( {\sum\limits_{p = 1}^{{N_t}} {{W_{n,p}}} {W_{{n_0},p}}} \right)} \!\!\right]} }}{{{N_t}{Z_{{n_0}}}(k)}}\!\!+\!\! {\tilde N_{\text{eff}}},\!\!
\end{equation}
where $\tilde{N}_{\text{eff}}$ is the equivalent noise.

\Needspace{5\baselineskip}
According to the row-orthogonality property of the Walsh-Hadamard matrix, $\sum_{p=1}^{N_t} W_{n,p}W_{n_0,p} = N_t \cdot \delta(n-n_0)$, where $\delta(\cdot)$ is the impulse function. This inner product property cancels out all cross-talk terms where $n \ne n_0$, resulting in $\hat{H}_{m,n_0}(k) \!=\! H_{m,n_0}(k) \!+\! \tilde{N}_{\text{eff}}$.

\rev{With SFD, a frequency-dependent phase rotation occurs between the left and right pilot estimates $\tilde{\mathbf{H}}_L$ and $\tilde{\mathbf{H}}_R$. For data symbol $d$ ($1\le d\le N_{\text{data}}$), $\tilde{\mathbf{H}}_k^{(d)}$ is reconstructed by linear interpolation between these bracketing pilots; the resulting chord error is quantified in Sec.~IV-C.}

For any valid subcarrier, the zero-forcing equalization completely eliminates spatial crosstalk, but it is extremely sensitive to both the condition number of the channel matrix and noise. In broadband LEO scenarios, high-frequency and non-uniform time-frequency coupling correlation inevitably causes some subbands to fall into areas of spatial rank deficiency. For the minimum mean square error (MMSE) equalizer defined as
\begin{equation}
\!\!\tilde{\mathbf{W}}_{\text{MMSE},k} \!=\! \left((\tilde{\mathbf{H}}_k^{(d)})^H \tilde{\mathbf{H}}_k^{(d)}\!+\! \frac{N_t}{\sigma^2} \mathbf{I}_{N_t}\right)^{-1} (\tilde{\mathbf{H}}_k^{(d)})^H,
\end{equation}
the matrix gain is actively reduced at frequencies with deteriorated condition numbers by introducing an adjustment factor, thereby \rev{suppressing the noise-amplification effect}.

\section{Performance Analysis}
\subsection{Channel Impairment and SINR Performance Loss}
The channel impairments can be classified into three main parts based on their sources: the free-space path loss, the spatial rank deficiency, and the residual clock deviation. This subsection quantifies the theoretical degradation of these impairments on the equivalent SINR.

\subsubsection{From Instantaneous Geometry to Condition Number}
To calculate the spatial degrees of freedom of the system, a conjugate symmetric matrix ${{\bf{R}}_H}(t,{f_k}) = {\bf{H}}_0^H(t,{f_k}){{\bf{H}}_0}(t,{f_k}) \in \mathbb{C}{^{{N_t} \times {N_t}}}$ is constructed. Its $(n,l)$-th element is the spatial inner product of the $n$-th and $l$-th transmit antennas:
\begin{equation}
\!\!\![\mathbf{R}_H(t,f_k)]_{n,l} \!=\! \sum_{m=1}^{M_r} \exp\left\{-j2\pi(f_c+f_k)(\tau_{m,n}(t)\!-\!\tau_{m,l}(t))\right\}\!.\!
\end{equation}
Its characteristic polynomial is
\begin{equation}
\!\!\!|\mathbf{R}_H \!\!-\! \lambda \mathbf{I}_{N_t}| \!=\! \lambda^{N_t} \!- a_{N_t-1}\lambda^{N_t-1} \!+\cdots - a_1\lambda \!+ a_0 = 0.\!
\end{equation}
The polynomial coefficients are determined by the trace and principal minors of the matrix varying with $(t,f_k)$. Since the ${\bf{R}}_H$ is a positive semi-definite matrix, this quartic equation must have some non-negative real roots $\lambda_1 \ge \lambda_2 \ge \cdots \ge\lambda_{N_t} \ge 0$ in the real domain. Matrix algebra root-finding gives $\kappa_{N_t\times N_t}(t,f_k) = \sqrt{\lambda_1(t,f_k)/\lambda_{N_t}(t,f_k)}$.

\subsubsection{Equalization SINR Loss}
For the MMSE equalizer, its single-frequency error covariance matrix is $(\mathbf{R}_H(t,f_k) + \sigma^2 \mathbf{I}_{N_t})^{-1}$.
Utilizing the algebraic determinant properties of the inverse matrix and adjugate matrix, the $i$-th element on the main diagonal of the error covariance matrix can be represented by its algebraic cofactor. The equivalent SINR for the signals transmitted from the $i$-th antenna is
\Needspace{7\baselineskip}
\begin{equation}
\mathcal{S}_{N_t \times N_t}^{(i)}(t,f_k) = \frac{|\mathbf{R}_H(t,f_k) + \sigma^2 \mathbf{I}_{N_t}|}{\sigma^2|\mathbf{R}_{H,\overline{i}}(t,f_k) + \sigma^2 \mathbf{I}_{N_t-1}|} - 1,
\end{equation}
where $\mathbf{R}_{H,\overline{i}} \in \mathbb{C}^{(N_t-1) \times (N_t-1)}$ is the principal submatrix after removing the $i$-th row and $i$-th column of the original matrix. The system-level average effective SINR is obtained through the average inverse mapping of all the MSE
\begin{equation}
\mathcal{S}_{cond}^{avg}(t,f_k) = \left( \frac{1}{N_t} \sum_{i=1}^{N_t} \frac{\sigma^2|\mathbf{R}_{H,\overline{i}} + \sigma^2 \mathbf{I}_{N_t-1}|}{|\mathbf{R}_H + \sigma^2 \mathbf{I}_{N_t}|} \right)^{-1} \!-\! 1.
\end{equation}

\subsubsection{The Chord Error by STO}
When performing linear interpolation using the left and right pilots separated by $\Delta T$, STO introduces a delay phase accumulation $\Phi_k = 2\pi f_k \varpi_m \Delta T$ that grows linearly with $f_k$.

At a relative position $\alpha \in (0,1)$, the true channel trajectory experienced by the data symbol is the complex plane arc ${\mathbf{H}}_{true}(\alpha) = {\mathbf{H}}_L e^{-j\alpha\Phi_k}$, while the mathematical trajectory of linear interpolation is the secant ${\mathbf{H}}_{intp}(\alpha) = (1\!-\! \alpha){\mathbf{H}}_L \!+\!  \alpha {\mathbf{H}}_L e^{-j\Phi_k}$. The resulting channel estimation error constitutes a multiplicative distortion $\vartheta(\alpha, \Phi_k) \!=\!  e^{-j\alpha\Phi_k} \!-\! (1\!-\! \alpha+\alpha e^{-j\Phi_k})$. Within the normal tracking range $|\Phi_k| \ll \pi$, performing a second-order Taylor expansion, it simplifies to
{\small
\begin{equation}
\!\vartheta(\alpha,\Phi_k) \!\approx\!\! \left(1\!-\!j\alpha\Phi_k\!-\!\frac{\alpha^2\Phi_k^2}{2}\right) \!\!-\! (1\!-\!\alpha) \!-\! \alpha\left(1\!-\!j\Phi_k\!-\!\frac{\Phi_k^2}{2}\right).
\end{equation}
}
Its corresponding chord error energy is approximated as the squared modulus of the complex difference as follows
\begin{equation}
\!\!\!\mathcal{E}_{chord}(\alpha,f_k) \!=\!{\left| {\vartheta (\alpha ,{\Phi _k})} \right|^2}\!\approx\! \frac{1}{4}(2\pi f_k\varpi_m\Delta T)^4(\alpha\!-\alpha^2)^2\!.\!
\end{equation}
Mathematically integrating above, the equivalent SINR for an arbitrary data symbol position and a specific subcarrier $f_k$ can be denoted as
{\small
\begin{equation}
\mathcal{S}_{final}(\alpha,t,f_k) \!=\! \left(\left(\mathcal{S}_{cond}(t,f_k)\!+\!1\right)^{-1} \!+\! \mathcal{E}_{chord}(\alpha,f_k)\right)^{-1} \!\!\!-\! 1.\!
\label{S_final}
\end{equation}
}

\subsection{Optimization Algorithm for Maximizing ECT}\label{sec4.3}

For a retained subcarrier $k$, the effective SINR $\gamma_k(t)$ gives the symbol error rate $P_s(\gamma_k)$ \cite{s4}. With Gray mapping, $P_b(\gamma_k) \approx P_s(\gamma_k)/\log_2(M)$. Independent FD demodulation gives $\gamma_k(t)=\mathcal{S}_{final}(\alpha,t,f_k)$ and the global uncoded BER $P_{b,\rm{unc}}(t) = \sum_{k \in \Omega_v} P_b(\gamma_k(t)) / N_v$. \rev{Substituting this BER into the decoder transfer curve gives the ECT for each candidate configuration.}

\SetKwInput{KwInput}{\textbf{Require}}
\SetKwInput{KwOutput}{\textbf{Output}}
\SetKwFor{For}{for}{do}{end}
\SetKwIF{If}{ElseIf}{Else}{if}{then}{else if}{else}{end}

\begin{algorithm}[t]
\small
    \caption{\rev{Joint truncation and VCM exhaustive search for maximizing dynamic ECT}}
    \KwInput{
        \rev{geometric parameters}, $t$, $R_{sym}$, $N_{\rm{FFT}}$,
        $\sigma^2$, $\varpi_m$;
        \rev{$\overline{\mathcal{M}} \!=\! \{4,16,64,256\}$ and 
        $\mathcal{K} \!=\! \{5,10,\ldots,50\}$};\!\\
    }
    \KwOutput{
        Optimal $M^*$, $\kappa_{th}^*$ and maximum $C_{eq}^*$.
    }

    Initialize $C_{eq}^* \gets 0$, $M^* \gets \text{null}$, $\kappa_{th}^* \gets \text{null}$\;
    
    Calculate $\mathbf{H}_0(t, f_k)$ and $\Delta\tau(t)$\;

    Calculate $\mathcal{S}_{{chord}}(f_k)$ and $\kappa(t, f_k)$ for all subcarriers\;
    
    \For{$\kappa_{th} \in \mathcal{K}$}{
        Calculate $\Omega_{v} = \{ f_k \mid \kappa(t, f_k) \le \kappa_{th} \}$ and $N_{v}$\;

       Calculate ${{\mathcal S}_{{{final}}}}(f_k)$ based on Eq.~\eqref{S_final} for all $k \in \Omega_{v}$\;

        \For{$M \in \overline{\mathcal{M}}$}{
            Calculate $P_{b,\rm{unc}}$ based on $M$ and $\gamma_{k}$\;

            Calculate $C_{eq}$ under current $(M, \kappa_{th})$\;
                
            \If{$C_{eq} > C_{eq}^*$}{
                    Update optimal state: $C_{eq}^* \gets C_{eq}$, $M^* \gets M$, $\kappa_{th}^* \gets \kappa_{th}$\;

            }
        }
    }
    \textbf{return} $M^*$, $\kappa_{th}^*$, $C_{eq}^*$\;
\end{algorithm}
\setlength{\textfloatsep}{0.5em}

\rev{Full-band high-order modulation degrades the global BER when broadband LoS-MIMO encounters strong time-frequency coupling, and the resulting BLER can reduce the ECT sharply. Subband truncation and variable coding and modulation (VCM) jointly regulate the modulation order $M$ and truncation threshold $\kappa_{th}$, balancing retained bandwidth against reliability.}

\rev{The discrete modulation order and the discontinuous dependence of $\Omega_v$ on $\kappa_{th}$ form a finite mixed-integer search. Alg.~1 evaluates the square-QAM set $\{4,16,64,256\}$ and $\kappa_{th}\in\{5,10,\ldots,50\}$ using the analytical SINR and ECT expressions. This range covers the MMSE operating region in which the retained bandwidth changes appreciably; the selected pair is stored as an ephemeris-indexed scheduling entry.}

In actual systems, the optimal decisions can be pre-bound with satellite-to-ground positions and stored in the look-up table of payload, realizing real-time optimal resource scheduling for dynamic orbits and maintaining optimal throughput continuously throughout the entire satellite overpass duration.

\subsection{Complexity Analysis and Implementability Evaluation}
\rev{The online computational load is counted per input sample, with block operations such as the FFT amortized over one OFDM symbol. The resulting real multiplication (RM), real addition (RA), and real division (RD) counts are summarized in Table~\ref{table_source}.}

\begin{table}[h]
\centering
\caption{Computational complexity of core modules.}
\renewcommand{\arraystretch}{1.1}
\setlength{\tabcolsep}{5.2pt}  
\begin{tabular}{|p{2.4cm}|c|c|c|c|c|c|}
\hline
\multirow{2}{*}{\textbf{Core Module}} & \multicolumn{2}{c|}{\textbf{RM}} & \multicolumn{2}{c|}{\textbf{RA}} & \multicolumn{2}{c|}{\textbf{RD}} \\ \cline{2-7} 
 & \textbf{2x2} & \textbf{4x4} & \textbf{2x2} & \textbf{4x4} & \textbf{2x2} & \textbf{4x4} \\ \hline
CFO compensate & 6 & 12 & 10 & 20 & 1 & 1 \\ \hline
Broadband FFT & 29.1/s & 58.2/s & 48.5/s & 97.0/s & 0 & 0 \\ \hline
Channel estimation & 12 & 48 & 28 & 176 & 0 & 0 \\ \hline
MMSE equalization & 85 & 692 & 141 & 1396 & 3 & 8 \\ \hline
\multicolumn{7}{p{8.4cm}}{
\rule{0pt}{1.0em}
Notation: the '/s' indicates that the resources of this module can be distributed across the entire OFDM symbol.
} \\ 
\end{tabular}
\label{table_source}
\vspace{-0.1em}
\end{table}

\rev{Following the operation-count convention in the implementable baseband analysis, one complex multiplication is represented by 3 RMs and 5 RAs, and one complex addition by 2 RAs. Channel estimation includes Walsh sign changes, accumulation over the pilot symbols, and one precomputed ZC factor per channel. The MMSE count comprises Hermitian Gram formation, diagonal loading, matrix inversion, weight extraction, signal recovery, and diagonal gain accumulation, while the FFT cost is amortized over one OFDM symbol. Table~\ref{table_source} therefore characterizes the online arithmetic load of the core modules.}

\section{Simulation and Performance Evaluation}

This section verifies the robustness of the proposed subband-adaptive transceiver baseband architecture under multi-dimensional constrained conditions through full-link system-level simulations, and evaluates the joint optimization gain of the dynamic ECT. Based on typical LEO constellation configurations, \rev{we establish a physical-layer system-level simulation model.} The system simulates a complete visible cycle of an LEO satellite traversing multiple distributed ground feeder receiving stations. The specific orbital and physical layer baseband parameter settings are detailed in Table~\ref{table_para}.

\begin{table}[h]
\centering
\caption{\small System and physical layer simulation parameters.}
\renewcommand{\arraystretch}{1.1}
\setlength{\tabcolsep}{6pt}  
\begin{tabular}{|p{1.5cm}|l|l|}
\hline
\textbf{Category} & \textbf{Parameter} & \textbf{Value} \\
\hline
\multirow{4}{=}{Geometric Parameters} 
  & Orbital altitude & \rev{$D=500$~km} \\\cline{2-3}
  & Satellite antenna spacing & \rev{$d_s=1.0$~m} \\\cline{2-3}
  & Ground station distance & \rev{$d_g=10$~km} \\\cline{2-3}
  & Geocentric angle  & $\theta = -15^\circ \!\sim\! 15^\circ$ \\\hline
\multirow{6}{=}{System Parameters} 
  & MIMO configuration & \rev{$2\times2$ / $4\times4$}\\\cline{2-3}
  & Carrier frequency & \rev{$f_c=73.5$~GHz} \\\cline{2-3}
  & Symbol rate & \rev{$R_{sym}=3$~Gsym/s} \\\cline{2-3}
  & FFT size  & $N_{\text{FFT}}=1024$ \\\cline{2-3}
  & CP length & $N_{\text{CP}}=32$ \\\cline{2-3}
  & Pilot/data symbols & \rev{$N_{\text{pilot}}/N_{\text{data}}=4/20$} \\\cline{2-3}
\hline
\end{tabular}
\label{table_para}
\vspace{-0.8em}
\end{table}

\subsection{Dynamic Degradation and SINR Bound Verification}

\rev{The analytical-bound check uses perfect CSI and ephemeris-aided NCO removal of the deterministic bulk Doppler, thereby isolating the spatial condition number and SFD-induced chord error.}

\begin{figure}[t]
\centering
\includegraphics[width=0.99\columnwidth]{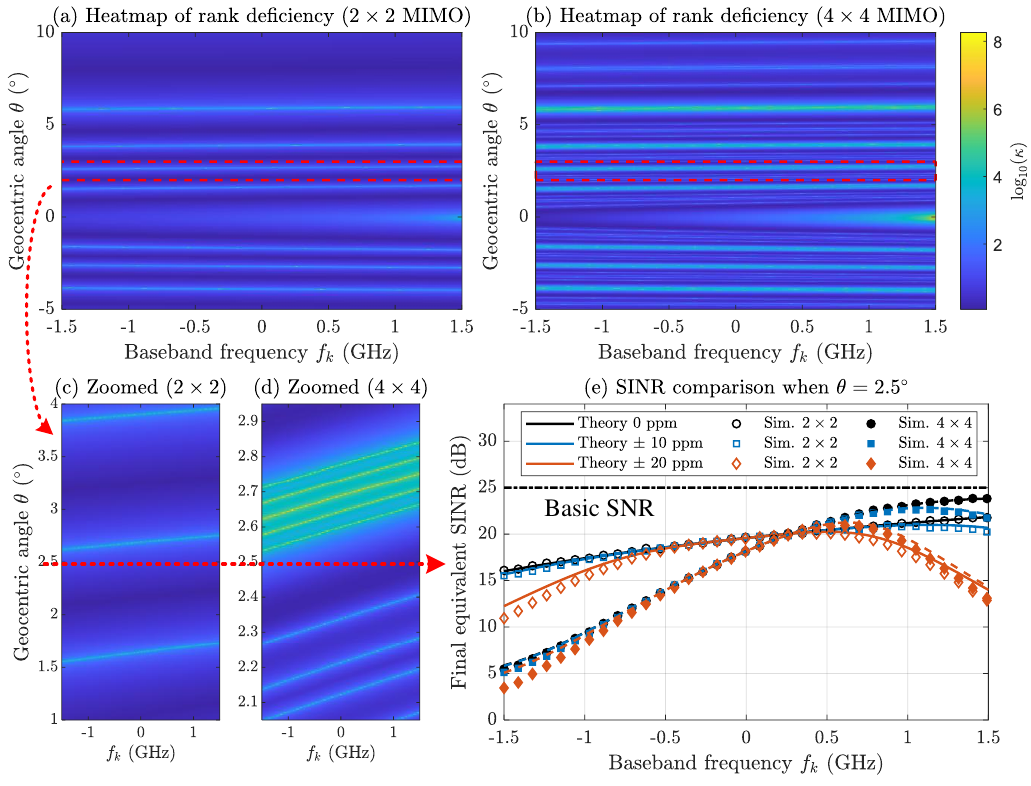}
\caption{Dynamic degradation trend and equivalent SINR.}
\label{fig2}
\vspace{0.5em}
\end{figure}

Fig.~\ref{fig2}(a) illustrates the thermal distribution of the condition number $\kappa(t,f_k)$ across the geocentric angle $\theta \in [-5^\circ, 10^\circ]$, while Fig.~\ref{fig2}(b) zooms in around $\theta = 2.5^\circ$. The time-frequency coupling correlation induces a steep spatial rank deficiency notch across the entire band. This notch exhibits diagonal sliding characteristics, indicating drastic variations with time and frequency.
\rev{Fig.~\ref{fig2}(c) compares the analytical SINR and Monte Carlo results at $\theta=2.5^\circ$ for oscillator deviations of 0, $\pm10$~ppm, and $\pm20$~ppm. The scatter points are obtained from complete IFFT/FFT processing and MMSE equalization. Away from the rank-deficient band, the simulated SINR approaches the 25~dB bound; inside the notch, it follows the condition-number-induced decrease. SFD introduces an additional symmetric loss because $P_{chord}\propto(f_k\sigma_m)^4$. At 1.6~GHz, the $\pm10$~ppm and $\pm20$~ppm cases lose approximately 2~dB and 10~dB, respectively. The agreement between the curves and scatter points validates the transition from the spatial/SFD model to the equivalent SINR in Section~IV.}

\subsection{BER Verification and Dynamic ECT Evaluation}

{
\begin{figure*}[t]
    \centering
    \begin{minipage}{0.32\textwidth}
        \centering
        \includegraphics[width=\textwidth]{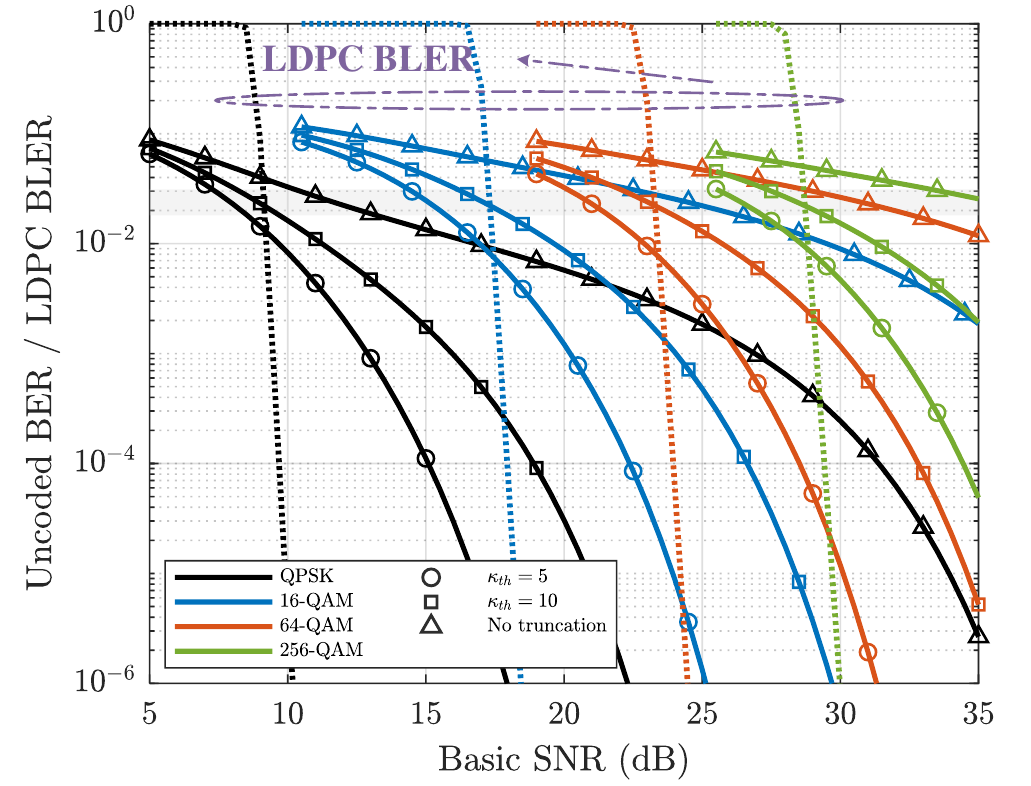}
        \caption{BER and LDPC BLER under modulation and truncation.}
        \label{fig3}
    \end{minipage}\hfill
    \begin{minipage}{0.32\textwidth}
        \centering
        \includegraphics[width=\textwidth]{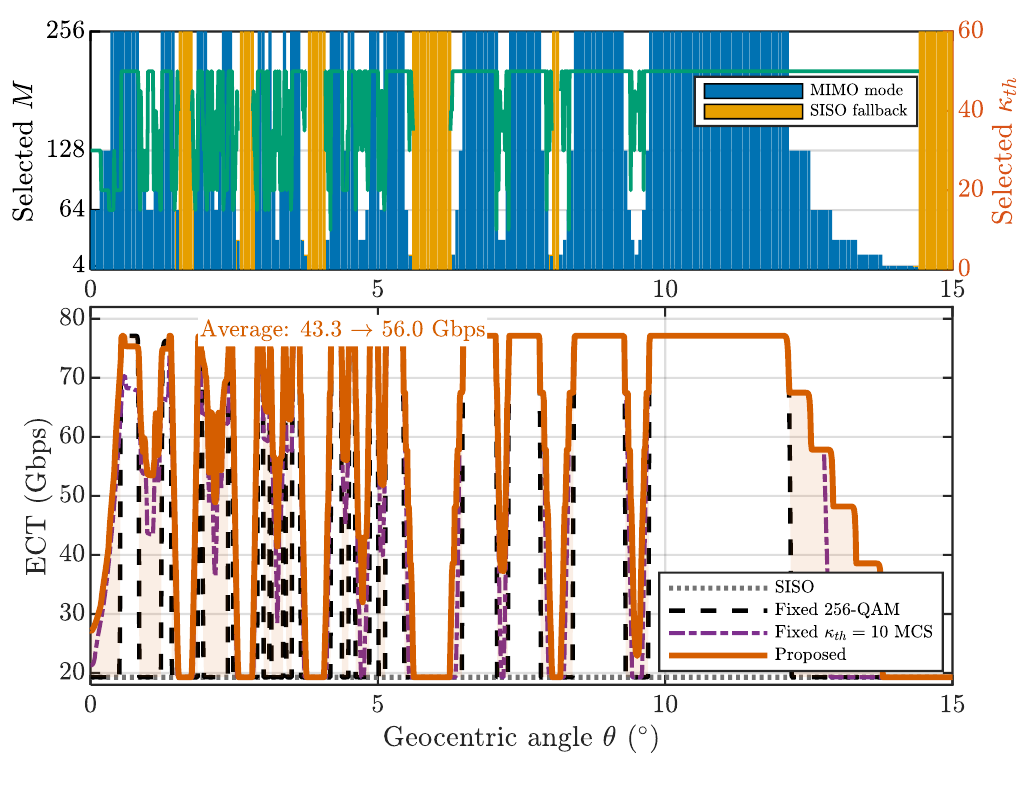}
        \caption{Dynamic scheduling and ECT with fixed-threshold and SISO baselines.}
        \label{fig4}
    \end{minipage}\hfill
    \begin{minipage}{0.32\textwidth}
        \centering
        \includegraphics[width=\textwidth]{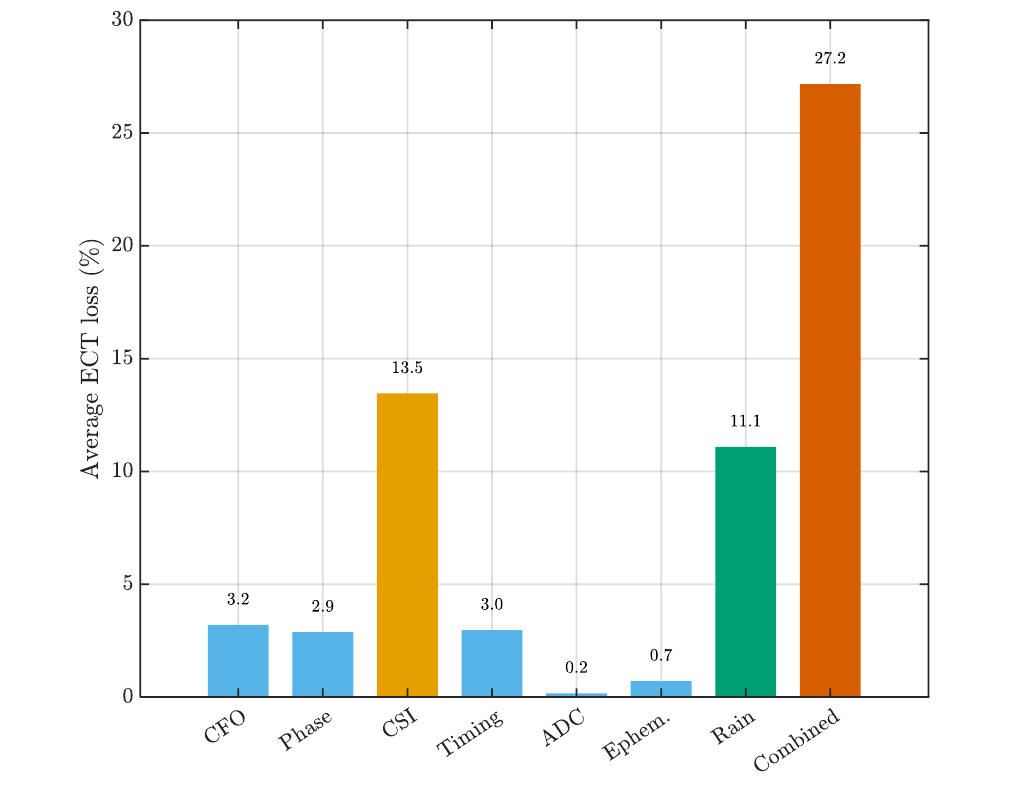}
        \caption{Average ECT loss from channel and receiver impairments.}
        \label{fig5}
    \end{minipage}
\vspace{1.0em}
\end{figure*}}

\rev{Fig.~\ref{fig3} compares QPSK, 16-QAM, 64-QAM, and 256-QAM at $\theta=2.5^\circ$ under $\kappa_{th}=5$, $\kappa_{th}=10$, and full-band transmission. Colors identify the modulation order, while circles, squares, and triangles identify the three truncation settings. Solid curves give the uncoded BER and dotted curves give the CCSDS (8160,7136) C2 LDPC BLER obtained from the fitted decoder characteristic. The curves are displayed from 5, 10.5, 19, and 25.5~dB for the four modulation orders to separate their useful transition regions. For a fixed modulation, reducing $\kappa_{th}$ rejects ill-conditioned subcarriers and lowers both the uncoded BER and the decoder waterfall SNR. The separation among the three marker sets is largest around the rank-deficient subbands; it narrows when their retained subcarrier sets become similar. For a fixed threshold, increasing the QAM order produces the expected rightward shift. The shaded interval marks the sampled pre-FEC BER range and supplies the continuous BLER mapping used in the ECT calculation.}

\rev{Fig.~\ref{fig4} evaluates Alg.~1 at a basic SNR of 30~dB and a residual oscillator deviation of 2~ppm. The upper panel shows the selected modulation order and truncation threshold: 256-QAM and a relaxed threshold are favored in well-conditioned intervals, whereas the scheduler reduces $M$, tightens $\kappa_{th}$, or selects one-stream transmission near a spatial notch. The lower panel compares full-band 256-QAM MIMO, fixed-$\kappa_{th}=10$ adaptive MCS, and the joint scheduler. The SISO curve remains at 19.28~Gbps because it is independent of the MIMO condition number and supplies the one-stream operating mode for all three schemes. Fixed-threshold MCS improves on fixed 256-QAM by changing $M$, but it cannot recover subcarriers when a looser threshold becomes feasible or reject additional tones at a deeper notch. The joint scheduler therefore follows the same high-ECT plateaus in well-conditioned intervals and separates from the fixed-threshold curve mainly during notch entry and recovery. Over $0^\circ$--$15^\circ$, the average ECTs of fixed 256-QAM, fixed-threshold MCS, and Alg.~1 are 43.30, 51.47, and 55.98~Gbps, respectively. Jointly adapting $M$ and $\kappa_{th}$ therefore yields an 8.8\% average gain over fixed-threshold MCS, with the largest improvement appearing around the moving rank-deficiency notches.}

\rev{Fig.~\ref{fig5} measures the average ECT sensitivity to receiver and propagation impairments under the same orbital sweep. Residual CFO, phase noise, timing error, an ADC resolution of 8~bits, and a 1~km ephemeris error produce losses of 3.2\%, 2.9\%, 3.0\%, 0.2\%, and 0.7\%, respectively. Among the receiver terms, CSI accuracy has the strongest effect: a CSI NMSE of $-25$~dB causes a 13.5\% loss, while 3~dB rain attenuation produces an 11.1\% loss. The nominal average ECT is 46.24~Gbps for this sensitivity sweep. With all modeled impairments enabled, it decreases to 33.68~Gbps, corresponding to a 27.2\% reduction. Because the decoder transfer and scheduling decisions are nonlinear, the combined loss is not the arithmetic sum of the individual bars. These link-level results characterize a single polarization; dual-polarization coupling, spatially nonuniform weather, and network scheduling can be incorporated in a system-level extension.}

\section{Conclusion} \label{sec5}
\rev{This paper analyzes the performance degradation caused by spatial rank deficiency and distributed clock deviation in broadband LEO LoS-MIMO systems. The proposed ECT model combines waveform overhead, equivalent SINR, and the CCSDS LDPC waterfall characteristic. A hardware-friendly subband-adaptive transceiver and joint VCM/truncation scheduler then bridge the moving spatial notches by coordinating the retained bandwidth, modulation order, and one-stream mode. The analytical bounds agree with the link-level simulations, and the scheduler achieves an average single-polarization ECT of 55.98~Gbps over the evaluated interval. These results provide a practical baseband approach for high-throughput LEO downlinks and a basis for subsequent dual-polarization and hardware implementation studies.}

\end{document}